\documentclass[aps,pra,reprint,superscriptaddress]{revtex4-2}

\usepackage{graphicx}
\usepackage{color}
\usepackage{amsmath,amssymb}
\usepackage{bm}
\usepackage{upgreek}
\usepackage{xspace}
\usepackage[colorlinks,urlcolor=blue,citecolor=blue,linkcolor=blue]{hyperref}
\usepackage{bbold}

\renewcommand{\selectlanguage}{}

\begin{document}

\title{Anisotropic electron-nuclear interactions in a rotating quantum spin bath}

\author{A.~A.~Wood}
\email{alexander.wood@unimelb.edu.au}
\affiliation{School of Physics, University of Melbourne, Parkville Victoria 3010, Australia}
\author{R. M. Goldblatt}
\affiliation{School of Physics, University of Melbourne, Parkville Victoria 3010, Australia}
\author{R. P. Anderson}
\affiliation{La Trobe Institute of Molecular Science, La Trobe University, Bendigo, Victoria 3550, Australia}
\author{L. C. L. Hollenberg}
\affiliation{School of Physics, University of Melbourne, Parkville Victoria 3010, Australia}
\affiliation{Centre for Quantum Computation and Communication Technology, University of Melbourne, Victoria 3010, Australia}
\author{R. E. Scholten}
\affiliation{School of Physics, University of Melbourne, Parkville Victoria 3010, Australia}
\author{A. M. Martin}

\affiliation{School of Physics, University of Melbourne, Parkville Victoria 3010, Australia}

\date{\today}
\begin{abstract}
The interaction between a central qubit spin and a surrounding bath of spins is critical to spin-based solid state quantum sensing and quantum information processing. Spin-bath interactions are typically strongly anisotropic, and rapid physical rotation has long been used in solid-state nuclear magnetic resonance to simulate motional averaging of anisotropic interactions, such as dipolar coupling between nuclear spins. Here, we show that the interaction between electron spins of nitrogen-vacancy centers and a bath of $^{13}$C nuclear spins in a diamond rotated at up to 300,000\,rpm introduces decoherence into the system via frequency-modulation of the nuclear spin Larmor precession. The presence of an off-axis magnetic field necessary for averaging of the dipolar coupling leads to a rotational dependence of the electron-nuclear hyperfine interaction, which cannot be averaged out with experimentally achievable rotation speeds. Our findings offer new insights into the use of physical rotation for quantum control with implications for quantum systems having motional and rotational degrees of freedom that are not fixed.  

\end{abstract}
\maketitle
An electron spin qubit and a surrounding bath of nuclear spins in diamond is a highly topical subject in solid-state quantum sensing and quantum information processing. The spin bath is both a deleterious source of decoherence, and a resource for storing and entangling quantum states. Understanding how the qubit-bath interaction behaves and can be controlled in a range of settings is an important consideration for future quantum technologies. For the nitrogen-vacancy (NV) centre in natural abundance diamond, the central electron spin is usually surrounded by a spin bath composed of spin-1/2 $^{13}$C nuclei. The nuclear spin bath is equally problematic for diamond-based quantum sensing~\cite{taylor_high-sensitivity_2008,rondin_magnetometry_2014, degen_quantum_2017} and quantum information processing~\cite{wrachtrup_processing_2006,prawer_quantum_2014}, since it leads to a noisy background magnetic field that limits the coherence time of the qubit and consequently the precision of quantum measurements~\cite{mizuochi_coherence_2009, balasubramanian_ultralong_2009, zhao_decoherence_2012, hall_analytic_2014}. Much effort has been devoted to dynamical-decoupling schemes to protect coherence and controllably isolate the central spin from the bath~\cite{lange_universal_2010, naydenov_dynamical_2011}.

The nuclear spins near optically-addressable electron spins like NV centers are also a resource for storing and retrieving quantum information, and form the basis of hybrid-spin quantum information processors~\cite{dutt_quantum_2007,neumann_multipartite_2008, maurer_room-temperature_2012, taminiau_universal_2014,waldherr_quantum_2014, chen_optimisation_2020}. Nuclear spins are also useful ancillae for quantum sensing, where the qubit-nuclear spin interaction is harnessed for more precise sensing~\cite{goldstein_environment-assisted_2011,unden_quantum_2016, haberle_nuclear_2017, rosskopf_quantum_2017, liu_nanoscale_2019} or accessing new sensing modalities~\cite{jaskula_cross-sensor_2018, ledbetter_gyroscopes_2012,ajoy_stable_2012, wood_magnetic_2017}. Novel means of spin bath control are thus sought to tailor the environment around the qubit as needed, decoupling spins in the bath from each other and also from the central electron spin. 

Physical rotation has long been employed in nuclear magnetic resonance (NMR) experiments as a means of suppressing deleterious spin interactions that broaden NMR transitions. Physical rotation of solid  state NMR samples with the magnetic field oriented at the magic angle ($\theta_m =54.74^\circ$) to the rotation axis simulates motional averaging of anisotropic interactions, a ubiquitous technique known as magic-angle spinning (MAS)~\cite{andrew_nuclear_1958, andrew_magic_1981}. The NV-nuclear spin hyperfine interaction is an archetypal case of a spin subject to an anisotropic interaction Hamiltonian, much like the chemical shift anisotropy or nuclear dipolar coupling in NMR~\cite{schmidt-rohr_multidimensional_2012} or anisotropic Zeeman interactions encountered in electron paramagnetic resonance (EPR)~\cite{schlosseler_electron_1998}. 

In the NV system, the hallmark of the NV-$^{13}$C hyperfine interaction is periodic collapse and coherent revival of the NV spin-echo signal~\cite{childress_coherent_2006}. Homonuclear dipole-dipole interaction between nuclear spins is much weaker ($\sim$kHz) and drives spin-conserving flip-flop interactions, creating a noisy magnetic environment for the NV and a reduced spin coherence time~\cite{maze_electron_2008, zhao_decoherence_2012} and attenuates the coherent revivals. In principle, the latter can be eliminated by physical rotation at kHz frequencies in the presence of a magnetic field at the magic angle to the rotation axis, though just how this affects the NV-$^{13}$C hyperfine interaction is less clear.

In this work, we investigate the qubit-bath interaction in a diamond subjected to physical rotation at rates comparable to the electron spin coherence time, $T_2$, with an off-axis magnetic field to introduce anisotropic averaging. Recent experimental advances have enabled quantum measurement and control of rapidly-rotating NV centres~\cite{wood_magnetic_2017, wood_quantum_2018}, making rotational averaging of deleterious spin interactions in a quantum setting a feasible new avenue of study for the central spin problem in diamond.  We find that the NV - nuclear spin hyperfine interaction is strongly affected by the rotation of the diamond, inducing decoherence due to rotation-dependent frequency modulation of the nuclear spin Larmor precession. While the coherence of the NV ensemble in the presence of a $^{13}$C bath for varying magnetic field angles has been previously examined in Ref. \cite{stanwix_coherence_2010}, our work shows that physical rotation of the diamond modulates the NV-$^{13}$C interaction in a way that leads to much faster dephasing than in the stationary case. We show theoretically that small tilts of the magnetic field from the rotation axis are expected to induce coherent modulation of the spin-echo signal at the rotation period, but for larger tilt angles the spin-echo signal is washed out by incoherent averaging of hyperfine-augmented nuclear precession of $^{13}$C nuclei close to the NV. Any effects of magic-angle spinning on the nuclear spin bath are therefore obscured, and we conclude new strategies to eliminate the electron-nuclear hyperfine coupling, or much faster rotation speeds, are required to realise rotational decoupling in the NV system.

Our study is motivated by the prospects of eliminating nuclear homonuclear couplings, which could potentially yield decoherence times comparable to isotopically-purified $^{12}$C diamonds~\cite{balasubramanian_ultralong_2009} and form a new mechanism for tuning the qubit-bath interaction. In previous work, it was found that physical rotation at a rate $\boldsymbol{\Omega}$ imparts an effective magnetic pseudo-field $\boldsymbol{B}_\Omega = \boldsymbol{\Omega}/\gamma_n$ to the $^{13}$C spin bath, $\gamma_n$ the nuclear gyromagnetic ratio~\cite{wood_magnetic_2017}. For the low magnetic fields that can be studied with NV experiments (compared to NMR fields exceeding $1\,$T), pseudo-fields can be made comparable to the bias magnetic field in the system, even to the point of eliminating altogether Larmor precession in the frame of the rotating NVs without any affect on the NV spin. The presence of the magnetic pseudo-field from physical rotation also has a significant effect in MAS, since the rotation itself generates a nontrivial field parallel to the rotation axis: for $f_\text{rot} = 5\,$kHz, $B_\Omega\sim 5\,$G. Consequently, the NV-$^{13}$C system represents a uniquely interesting system in which to study the low-field dynamics of motional averaging.

This paper is structured as follows. We review in Section \ref{sec:theor} the relevant theory concerning the NV-$^{13}$C interaction pertinent to our studies. Section \ref{sec:exp} describes the setup and our experimental examination of the loss of NV spin coherence for varying magnetic field angles and rotation speeds. We use a theoretical model of the NV coupled to a bath of interacting spins in Section~\ref{sec:theor2} to confirm our findings and examine situations too difficult to realise experimentally. In Section~\ref{sec:disc} we discuss our key findings in detail and give an outlook on rotation as a means of controlling spin coherence in future work.

\section{NV-nuclear spin interactions in the rotating frame}\label{sec:theor}
\begin{figure*}[t]
	\centering
		\includegraphics[width = \textwidth]{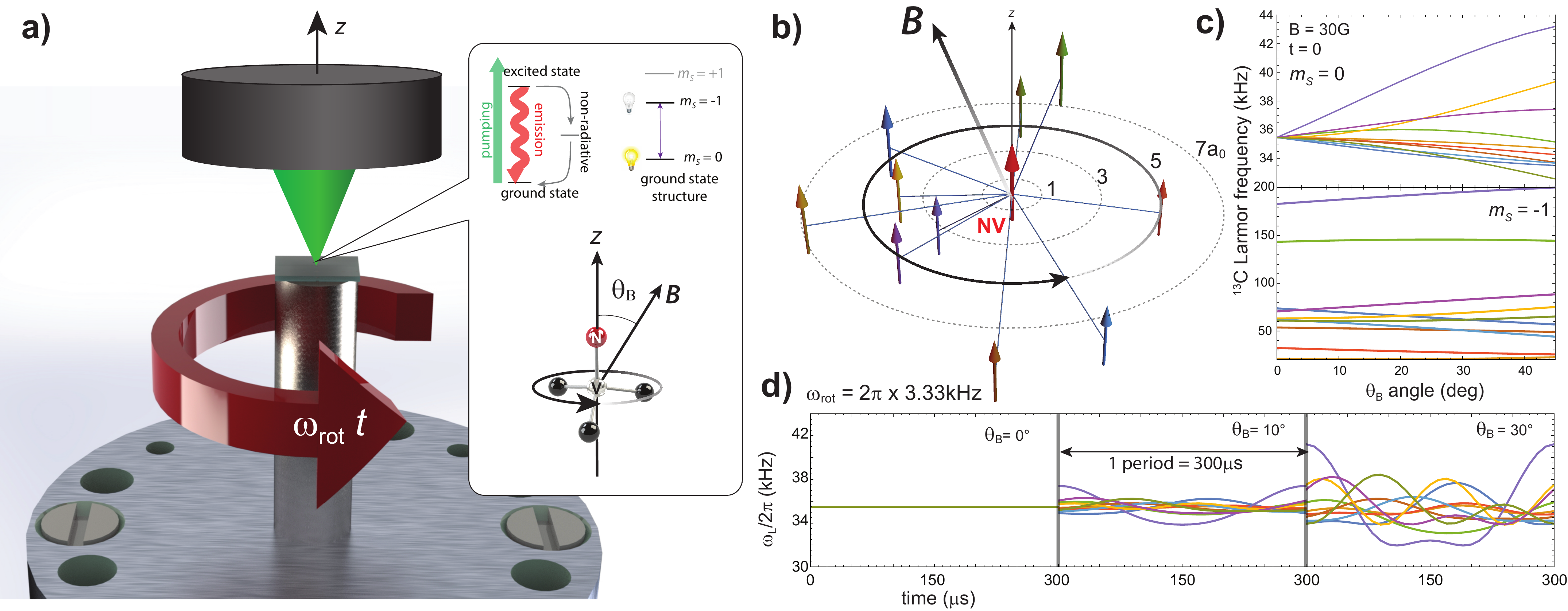}
	\caption{a) A diamond is mounted on an electric motor such that the $\langle111\rangle$ crystallographic direction is approximately parallel to the rotation axis. An ensemble of NV centers are optically prepared and readout by a confocal microscope, and magnetic fields are applied to tilt the total field $\boldsymbol{B}$ away from the rotation axis by an angle $\theta_B$ (inset). b) NV electron spin (red) interacting with a bath of 20 $^{13}$C nuclear spins within a diamond physically rotating around the $z$-axis at $\omega_\text{rot}$. The dashed circles depict radii in units of lattice constant $a_0 = 0.154$\,nm. c) Nuclear spin precession frequencies at $t=0$ for $|\boldsymbol{B}| = 30\,$G and varying tilt angles for the NV electron spin in the $m_S = 0$ state (top) and $m_S = -1$ state (bottom). Note the shift from the bare precession frequency of $32.1\,$kHz due to the rotation at $3.33\,$kHz. The curve color corresponds to the coloured vector in (b). d) Nuclear precession frequencies of the spin ensemble for $m_S = 0$, $\omega_\text{rot} = 3.33\,$kHz as a function of time for different tilt angles $\theta_B$. Rotation of the diamond modulates the precession frequency of the nuclear spins due to anisotropic, non-secular terms in the NV-$^{13}$C hyperfine interaction.}
	\label{fig:fig1}
\end{figure*}

A simplified schematic of our experimental configuration, important geometrical quantities and the NV energy level scheme is depicted in Fig.~\ref{fig:fig1}(a). The negatively-charged NV centre in diamond~\cite{doherty_nitrogen-vacancy_2013, schirhagl_nitrogen-vacancy_2014} features a spin-1 ground state with an $m_S = 0$ and degenerate $m_S = \pm1$ spin sublevels split by $D_\text{zfs} = 2870\,$MHz. Application of a magnetic field $\boldsymbol{B}$ breaks the degeneracy of the magnetic sublevels, leading to a splitting of $\omega_m \approx D_\text{zfs}+m \gamma_e |\boldsymbol{B}|\cos\theta_B$, where $\gamma_e/2\pi = 2.8\,$MHz/G is the electron gyromagnetic ratio and $\theta_B$ the angle between the NV axis and the magnetic field, an approximation valid for low magnetic fields ($<50$G) and $\theta_B\ll 90^\circ$. The diamond sample we consider in this work is natural abundance $^{13}$C CVD material ($98.9\%^{12}$C) and contains an ensemble density of NV centres, with a typical nitrogen density of $0.1\,$ppm. We also work at magnetic fields far from any cross-relaxation resonances between the NV and its intrinsic nuclear spin~\cite{jacques_dynamic_2009}, nearby P1 centers, and $^{13}$C spins~\cite{wunderlich_optically_2017,pagliero_multispin-assisted_2018}. The Hamiltonian for the NV-$^{13}$C system in the stationary laboratory frame is \cite{zhao_decoherence_2012} 
\begin{equation}
H = H_\text{NV} + H_\text{Z} + H_\text{hfs} + H_\text{d-d},
\label{eq:hintfull}
\end{equation}
where the NV Hamiltonian is $H_\text{NV} = D_\text{zfs} S_z^2$, $H_\text{Z} = \gamma_e \boldsymbol{B}\cdot\boldsymbol{S}+\gamma_{n} \boldsymbol{B}\cdot\boldsymbol{I}$ is the Zeeman Hamiltonian with $\gamma_n/2\pi = 1071.5$Hz/G the $^{13}$C gyromagnetic ratio, $H_\text{hfs} = \boldsymbol{S}\cdot\sum_i \boldsymbol{A}_i\cdot\boldsymbol{I}_i$ is the hyperfine interaction between the NV spin $\boldsymbol{S}$ and the $i$th nuclear spin $\boldsymbol{I}_i$, $\boldsymbol{A}_i$ the hyperfine tensor and $H_\text{d-d}$ represents the dipole-dipole interactions between $^{13}$C spins. 

The focus of our analysis is the hyperfine interaction tensor between the NV and nuclear spin, $\boldsymbol{A}_i$, which contains a contact term due to the finite extent of the electron wavefunction, and a dipole-dipole interaction term. The Fermi contact term scales exponentially with distance and thus dominates for nuclear spins within a few lattice sites of the NV. The dipolar term decreases with NV-nuclear spin separation as $r_i^{-3}$, and is responsible for most dynamics observed in a $\tau\sim T_2$ duration spin-echo experiment, such as coherent revivals at half the Larmor period~\cite{childress_coherent_2006}. The dipolar term is also responsible for the behaviour we observe as the magnetic field is tilted away from the rotation axis~\cite{stanwix_coherence_2010}.

Approximating the NV and $^{13}$C as point dipoles, the hyperfine interaction tensor is given by~\cite{zhao_decoherence_2012}
\begin{equation}
\boldsymbol{A}_i = \frac{\mu_0 \gamma_e \gamma_{n} \hbar}{4\pi r_i^3}\left(1-3\boldsymbol{\hat{r}}_i\boldsymbol{\hat{r}}_i\right),
\label{eq:hintfull2}
\end{equation}
where $\mu_0$ is the vacuum permittivity and $\boldsymbol{\hat{r}}_i$ a unit vector along the axis connecting the NV to the $i$th nuclear spin. The presence of magnetic field components that are not parallel to the NV axis (non-secular terms) has an important effect on the hyperfine interaction. First reported in Ref.~\cite{childress_coherent_2007} for single NV centers and later by Ref.~\cite{stanwix_coherence_2010} for NV ensembles, the off-axis magnetic field mixes electron spin states, encoding each nuclear spin with an $r_i$-dependent precession frequency that asymptotes to the bare precession frequency $\gamma_{n}B$ as $r_i$ increases. The increased spread of nuclear spin precession frequencies results in faster decoherence in a spin-echo experiment. As we shall see, the encoding of nuclear spin precession rates is not rotationally symmetric, meaning that azimuthal rotation of the diamond (or magnetic field) results in frequency modulation of the nuclear spins.

\begin{figure*}
	\centering
		\includegraphics[width = \textwidth]{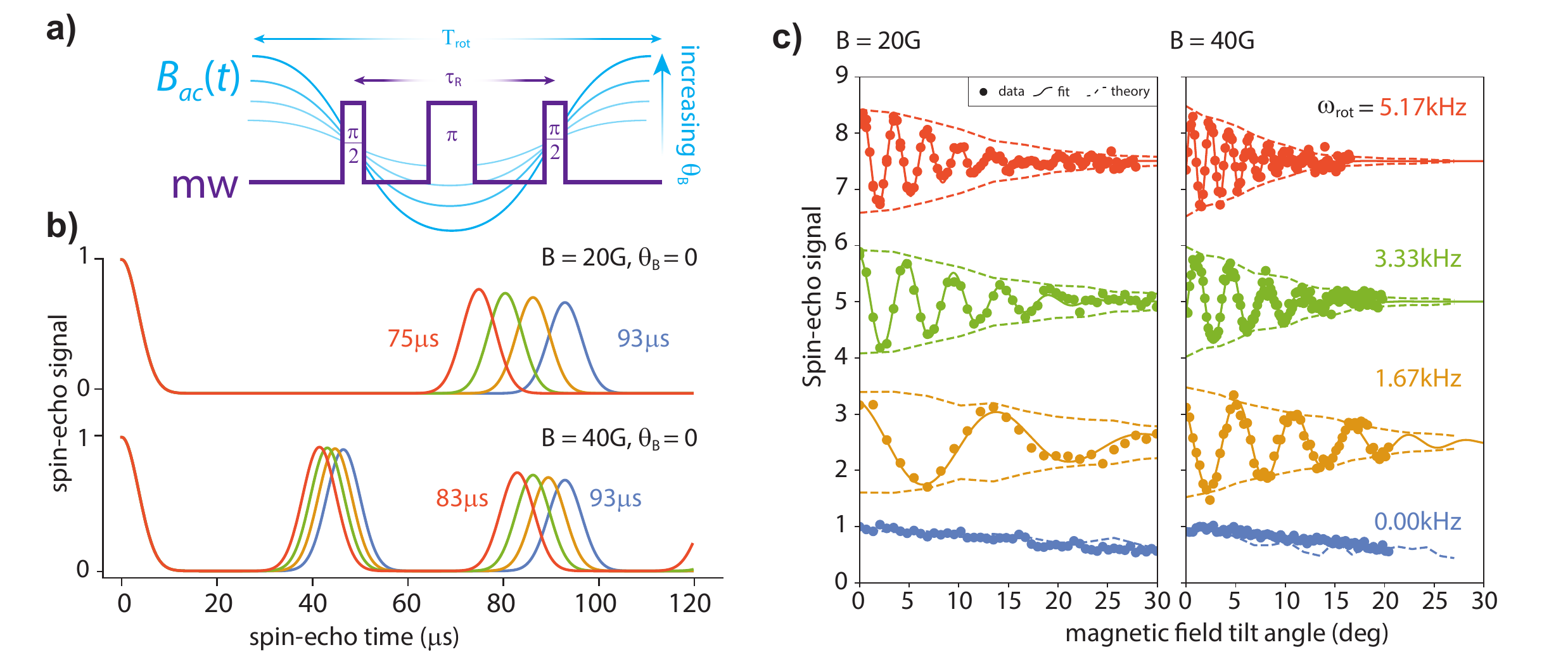}
	\caption{Measuring the coherence of a rapidly rotating NV - nuclear spin ensemble. (a) Rotating the diamond with the NV axis tilted slightly away from the rotation axis up-converts magnetic field components orthogonal to the rotation axis to fields oscillating at the rotation frequency. This modulated field can then be measured by a spin-echo pulse sequence applied synchronous with the rotation of the diamond. Tilting the magnetic field further from the rotation axis increases the amplitude of the up-converted field, and results in fringes in the spin-echo signal. (b) Illustrative time-domain spin-echo signals for $B = 20\,$G (top) and $B = 40\,$G (bottom) for $\theta_B = 0$ showing rotationally-induced shifting of the $^{13}$C contrast revivals. Adding a transverse field to change $\theta_B$ increases the total field, which also shifts the revival. We adjust the interpulse time $\tau_R$ so that the final $\pi/2$-pulse of the spin echo sequence occurs at the same time as first (20\,G) or second (40\,G) contrast revival maximum for a given magnetic field configuration. (c) Spin-echo signal at fixed $\tau_R$ vs. $\theta_B$ for different rotation speeds. An exponentially-damped sinusoid (solid line) is fitted to the experimental data (points), and a fully constrained theoretical model of the envelope is overlaid (dashed line). We observe a rapid reduction of contrast with $\theta_B$ as the rotation speed is increased that matches well with theory; the discrepancy is attributed primarily due to imperfect magnetic field calibration. Our observations are consistent with dephasing introduced by NV-$^{13}$C hyperfine modulation arising from rotation in the presence of an off-axis magnetic field.}
	\label{fig:fig2}
\end{figure*}

The Hamiltonian Eq.\ref{eq:hintfull} consists of terms intrinsic to the diamond, $H_\text{in} = H_\text{NV} + H_\text{hfs} + H_\text{d-d}$, and extrinsic lab-frame terms, $H_\text{ex} = H_Z$. Physical rotation of the diamond about the $z$-axis in space corresponds to application of a unitary transformation $\mathcal{R}(t) = \exp\left(-i J_z \omega_\text{rot} t\right)$, with $J_z  = I_z + S_z$, to either the intrinsic or extrinsic terms. Since the NV axis is very nearly parallel to the rotation axis in our experiments, there is no substantive difference between the two transformations, although it is marginally simpler to rotate the extrinsic Hamiltonian and consider the effects in the NV frame. The transformed Hamiltonian becomes

\begin{align}
H_\text{rot} & = H_\text{NV}  + H_\text{hfs} + H_\text{d-d} \nonumber \\
             & + \mathcal{R}(t)H_Z \mathcal{R}^\dagger(t) + i \frac{d}{dt}\mathcal{R}(t)\mathcal{R}^\dagger(t).
\label{eq:hrot}
\end{align}

The final term in Eq. \ref{eq:hrot}, $ i \frac{d}{dt}\mathcal{R}(t)\mathcal{R}^\dagger = \omega_\text{rot} J_z$, is responsible for the rotationally-induced magnetic pseudo-field shift~\cite{wood_magnetic_2017} of the $^{13}$C nuclear spin precession frequency. Diagonalizing the Hamiltonian Eq. \ref{eq:hrot} at time $t$ yields the eigenenergies $E^i_{m_S, m_I}$ for the $i$th nuclear spin, and computing 
\begin{equation}
\omega^i_L (m_S) = \frac{1}{\gamma_n}\left|E^i_{m_S, \uparrow}-E^i_{m_S, \downarrow}\right|
\label{eq:wl}
\end{equation}
yields the $m_S$-dependent Larmor frequency for the $i$-th nuclear spin. Figure \ref{fig:fig1}(b,c) shows the nuclear spin Larmor precession frequencies for $N$ nuclear spins surrounding the NV centre for varying magnetic field tilt angle $\theta_B$, and Fig. \ref{fig:fig1}(d) depicts the time-dependence of $\omega_L (m_S)$ for several tilt angles. The rotating magnetic field (in the frame of the NV) leads to frequency modulation of the nuclear spin precession in a manner dependent on the position of the nuclear spin. This modulation in turn introduces a form of ensemble dephasing, since the spread of nuclear spin precession frequencies will strongly suppress the observation of coherent behaviour, such as $^{13}$C spin-echo revivals.   

The non-secular terms in the NV-$^{13}$C hyperfine interaction are responsible for the modulation of the nuclear spin precession. A useful representation that offers insight into the role of non-secular terms is considering an $m_S$-dependent effective magnetic field acting on the nuclear spin~\cite{childress_coherent_2006,zhao_anomalous_2011, huang_observation_2011, reinhard_tuning_2012}, which also depends on the angle the magnetic field makes to the NV axis~\cite{maze_electron_2008,stanwix_coherence_2010}. Since the electron spin interacts much more strongly with magnetic fields than the nuclear spins, there is significant back-action on the bath~\cite{zhao_sensing_2012} and as a result the electron spin state determines the evolution of the coupled nuclear spins. The state-dependent field for the $i$th nuclear spin is given by  
\begin{equation}
\boldsymbol{B}_{\text{eff},i}(m_S) = m_S \boldsymbol{B}_\text{dip}+\boldsymbol{B}\cdot\boldsymbol{g}_i
\label{eq:beff}
\end{equation}
where $\boldsymbol{B}_\text{dip}$ is the electron spin dipole field and $\boldsymbol{g}_i$ is an effective nuclear spin $g$-tensor that expresses the NV-$^{13}$C hyperfine-mediated modification to the nuclear spin precession frequency
\begin{equation}
\boldsymbol{g}_i = \mathbb{1}-\frac{\gamma_e}{\gamma_{n} D_\text{zfs}}(2-3|m_S|)\boldsymbol{A}_i.
\label{eq:gfact}
\end{equation}
Equation \ref{eq:beff} concisely expresses that the $^{13}$C spin experiences a magnetic field depending on the NV electron spin state. The resulting Larmor precession rate of the nuclear spins can therefore be written as 
\begin{equation}
\omega^i_{L}(m_S) = \gamma_{n}\left|\boldsymbol{B}_{\text{eff},i}(m_S)\right|.
\label{eq:}
\end{equation}
When either the diamond or the magnetic field is rotated, the $\boldsymbol{B}\cdot\boldsymbol{g}_i$ term in Eq. \ref{eq:beff} leads to time dependence of $\omega^i_{L}$. In the following sections of this paper we will discuss experimental detection of this frequency modulation effect.

\section{Experiment}\label{sec:exp}
Our experimental setup is essentially the same as that described previously~\cite{wood_magnetic_2017}. An optical-grade CVD diamond is mounted on the spindle of an electric motor (Fig. \ref{fig:fig1}(a)) that can spin at up to $350,000$\,rpm. The diamond is a $(111)$-cut sample with a $\sim 4^\circ$ polish (miscut) angle, and the motor tip is machined so that one orientation class of NVs makes an angle of $<0.2^\circ$ to the rotation axis. A 6\,mm-working-distance microscope objective focuses $500\,\upmu$W of green light close to the rotation center of the spinning diamond, and collects emitted red fluorescence which is directed to an avalanche photodiode, forming a confocal microscope. Three pairs of current carrying coils are used to generate magnetic fields of up to $\approx40\,$G with a maximum tilt angle from the rotation axis of $\theta_B = 40-50^\circ$. We used vector magnetometry to calibrate the field produced from the coils, allowing us to estimate the tilt angle used in experiments. A 20\,$\upmu$m copper wire, 300$\upmu $m above the surface of the diamond, is used to apply microwave fields with typical Rabi frequencies of $5\,$MHz. 

We use spin-echo interferometry~\cite{hahn_spin_1950} to measure the coherence of the rotating ensemble of NV centers. A $3\,\upmu$s laser pulse is first applied, followed by microwave pulses synchronous with the rotation~\cite{wood_quantum_2018}. Although the NV orientation class we target is nearly parallel to the rotation axis, the small misalignment $\delta\theta$ between the NV axis and the $z$ rotation axis means that static magnetic field components $B_\perp$ transverse to $z$ are effectively up-converted to alternating fields at the motor rotation frequency, $B_\text{ac} \approx  B_\perp \delta\theta \cos(\omega_\text{rot}t-\phi_0)$~\cite{wood_$t_2$-limited_2018}, with $\phi_0$ an inconsequential phase set by the orientation of the diamond on the motor. Fluorescence detected from a spin-echo sequence with fixed total time $\tau$ will therefore exhibit fringes as the magnetic field tilt angle $\theta_B$ is increased. We use the amplitude of these fringes to infer the coherence of the NV electron spin as a function of magnetic field tilt angle and rotation speed of the diamond. The spin-echo time $\tau_R$ must be set to an integer multiple of twice the $^{13}$C nuclear spin precession frequency, where coherence revivals occur. These revivals shift significantly in time due to the increasing rotation speed~\cite{wood_magnetic_2017} and also slightly when a transverse field $B_\perp$ is added to vary the tilt angle. The value of $\tau_R$ is set to  $2 n (\gamma_n B_\text{tot}/2\pi + f_\text{rot})^{-1}$, for integer $n$ and $B_\text{tot} = (B_\perp^2+B_z^2)^{1/2}$ to ensure overlap with the coherence revival, as depicted in Fig.~\ref{fig:fig2}(a,b).  

Figure \ref{fig:fig2}(c) shows spin-echo fringes for increasing magnetic field tilt angles for two different magnetic field strengths, $20\,$G and $40\,$G, and rotation speeds ranging from stationary to $5.17\,$kHz. We set the spin-echo time to the first coherence revival for $20\,$G (at $\tau_R = 75-93\,\mu$s) and the second revival for $40\,$G (at $\tau_R = 83-93\,\mu$s) and change the tilt angle by applying a transverse magnetic field $B_\perp$ along the $x$-axis. The spin-echo sequence is synchronised with an arbitrary phase to the rotation, and hence up-converted magnetic field. This phase in turn sets the period of the resulting fringes, and is scaled with rotation speed to remain approximately constant. As the rotation speed increases, we observe a rapid damping of contrast for increasing magnetic field tilt angle. 

\section{Theoretical investigation}\label{sec:theor2}

\begin{figure}
	\centering
		\includegraphics[width = \columnwidth]{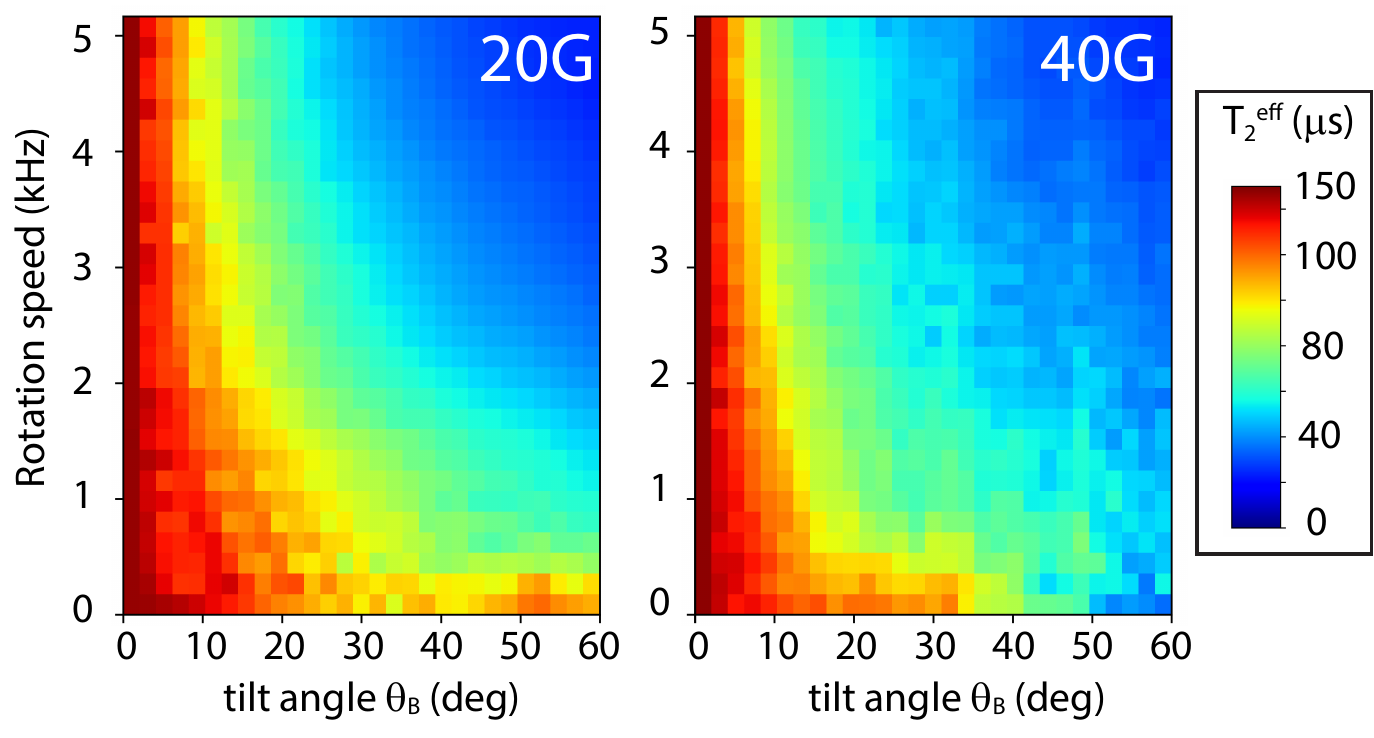}
	\caption{Simulated effective electron spin coherence time $T_2^\text{eff}$as a function of magnetic field tilt angle $\theta_B$ and rotation speed $\omega_\text{rot}$ for $B = 20\,$G (left) and 40\,G (right). Each value of $T_2$ derives from an exponential fit to the spin-echo signal envelope, evaluated at the rotationally-shifted $^{13}$C revival positions and averaged over 10 distinct bath configurations, each with approximately 125 $^{13}$C nuclei at random lattice sites. A decay envelope with time constant $T_2^{\text{phenom}}$ limits the maximum coherence time to $150\,\upmu$s to simulate the effects of other defects within the diamond sample used in this work.}
	\label{fig:fig3}
\end{figure}

We compared our results to a numerical simulation of the NV spin-echo signal generated by adapting the disjoint-cluster method of Ref. \cite{maze_electron_2008} to the time-dependent, non-secular case. The large ensemble of $N$ interacting spins is decomposed into $n$ groups $\left\{G_1, G_2, ..., G_n\right\}$ of up to $g$ spins, with strong interactions between elements of each group and negligible interactions between elements in different groups. Considering a single NV interacting with a group of up to $g$ $^{13}$C nuclei, the density matrix of the system within the group is given by $\rho = \rho_e\otimes\rho_B$, with $\rho_e$ the density matrix of the electron spin and $\rho_B = 1/2^g \mathbb{1}^g$ the density matrix for the thermally-populated nuclear spin bath. The time-evolution of $\rho$ is then 
\begin{equation}
\rho(t) = U_\tau\rho(0)U^\dagger_\tau,
\end{equation}
with $U_\tau = U_{\pi/2} U(\tau/2) U_\pi U(\tau/2)U_{\pi/2}$ a unitary operator representing a spin-echo sequence applied to the electron spin with free-evolution time $\tau$, $U_{\pi/2}, U_\pi$ operators for $\pi/2$ and $\pi$ pulses, $U(t)$ the free evolution operator and $\rho(0) = |0\rangle\langle0|\otimes\rho_B$. The normalised spin-echo signal for this particular group of nuclear spins $G$ is then given by
\begin{equation}
S_{G}(\tau) = 2\text{Tr}\left[\mathcal{P}_0\rho(\tau)\right]-1,
\end{equation}
with $\mathcal{P}_0$ the projection operator onto the $|m_S = 0\rangle$ electron spin state. The spin-echo signal from all groups is then $S(\tau) = \prod_G S_G(\tau)$. To simulate ensemble averaging due to multiple NV centres, we then compute the average signal $S_\text{ave} = 1/N_\text{ave}\sum S(\tau)$ for $N_\text{ave}$ different configurations of $^{13}$C baths. The simulation results depicted in Fig. \ref{fig:fig2}(c) are for $N_\text{ave} = 20$ different configurations of up to $N = 310$ $^{13}$C spins with a maximum group size of $g = 3$. Our selection of $g = 3$ includes spin correlations up to third order, but to include any effects of potential homonuclear dipolar decoupling (a two-body interaction) we need only use $g = 2$. Short experimentally realisable spin coherence times also prevent any higher order effects from manifesting (see below). 

\begin{figure}
	\centering
		\includegraphics[width = \columnwidth]{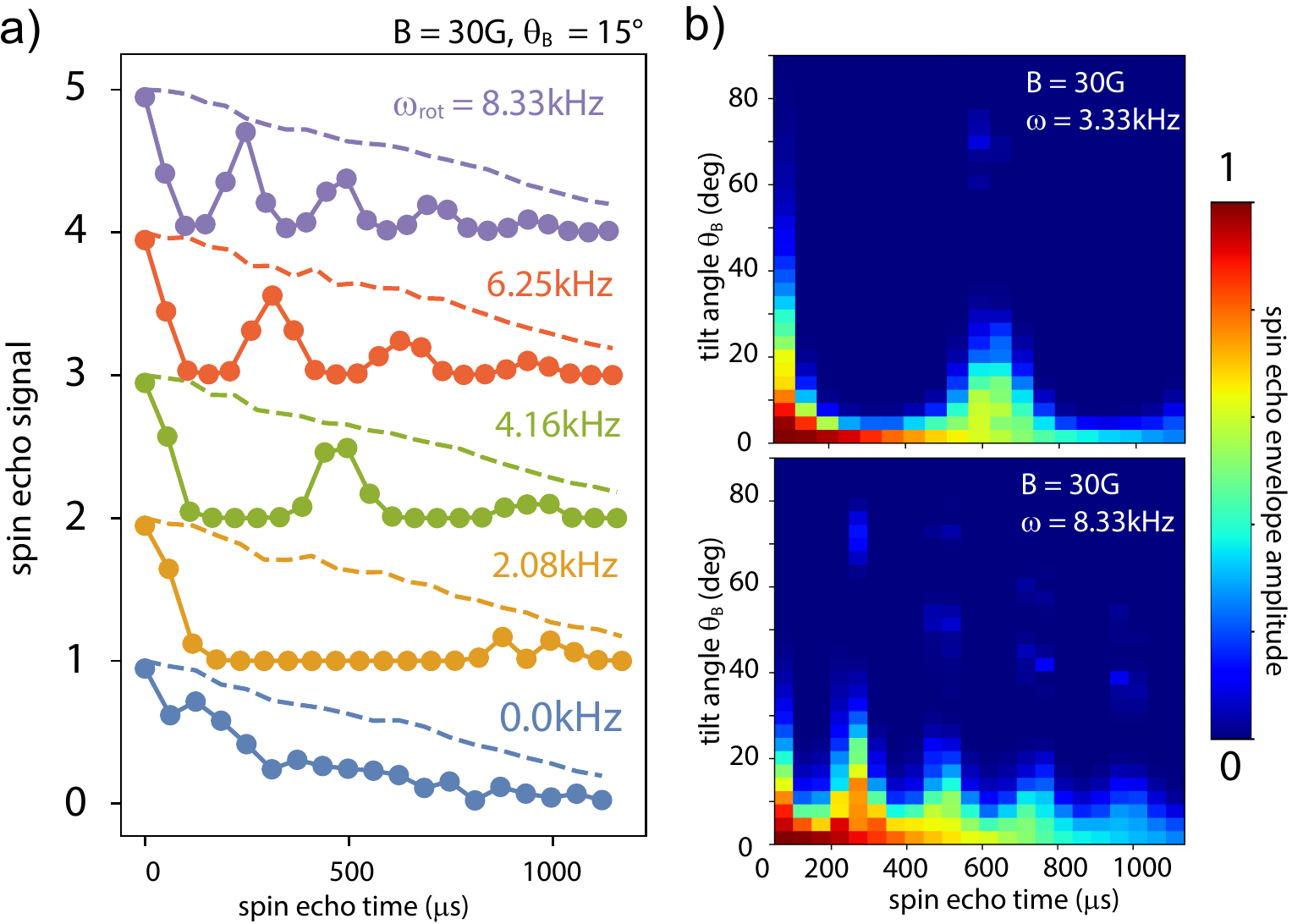}
	\caption{Simulating the onset of spin-echo envelope modulation due to rapid rotation with a tilted magnetic field. a) for small $\theta_B$ the spin-echo signal collapses as the rotation speed is increased, before contrast revivals at twice the rotation period become evident. The dashed traces depict the $^{13}$C dipole-dipole limited coherence envelope for the relevant rotation speed and $\theta_B = 15^\circ$. b) for fixed rotation speeds of 200\,krpm (3.33\,kHz, top) and 500\,krpm (8.33\,kHz, bottom), increasing the tilt angle causes the rotational revivals to decay. At larger angles, small, intermittent revivals reappear at integer multiples of the rotation period. Each simulated spin-echo trace derives from the average of at least 20 different bath configurations of up to 300 $^{13}$C nuclear spins. }
	\label{fig:fig4}
\end{figure}

The theory simulations show concordance with the experimental results and reproduce closely the dephasing induced by rapid rotation in the presence of an off-axis magnetic field. Since this effect is entirely due to NV-nuclear spin hyperfine coupling (i.e. $g = 1$), retaining the homonuclear coupling terms, or for that matter the grouping schema based on the homonuclear couplings, is redundant. We include these terms, and correlations up to $g = 3$ for the purposes of extending the model to situations where longer coherence times are possible, as detailed later in this work.

It is useful to consider a figure of merit which parametrizes the coherence of the system, such as an effective spin coherence time. We define the effective spin-coherence time $T_2^\text{eff}(\theta_B, \omega_\text{rot})$ as the $1/e$-damping time of the spin-echo contrast for a given tilt angle and rotation speed. We calculate the spin-echo signal at integer multiples of the rotationally-shifted $^{13}$C contrast revival time $\tau_R = 4\pi/(\gamma_{13}B + \omega_\text{rot})$, and, assuming the spin-echo signal damps as $S(\tau) = S(0)e^{-(\tau/T^\text{eff}_2(\theta_B, \omega_\text{rot}))^n}$, fit a stretched exponential decay model to extract $T^\text{eff}_2(\theta_B, \omega_\text{rot})$. We include also a phenomenological damping term to reproduce the stationary, $\theta_B = 0$ coherence time $T_2^{\text{phenom}} =  150\,\upmu$s, due to other defects in the diamond (principally paramagnetic nitrogen P1 centers~\cite{bauch_decoherence_2020}). In Fig. \ref{fig:fig3}, we evaluate $T_2^\text{eff}$ for variable $\theta_B$ and $\omega_\text{rot}$ for the magnetic field strengths examined in this work, $20$ and $40$\,G. As expected, we observe a rapid decay of spin coherence time with increasing rotation speed and tilt angle that becomes sharper as the magnetic field strength increases. Figure \ref{fig:fig3} broadly summarises the range of potential coherence times possible with our current experimental configuration.  

With our experimental results well explained by the theoretical model, we can confidently examine NV-$^{13}$C dynamics outside the domain of measurement times, magnetic fields, tilt angles and rotation speeds accessible with our current experimental setup. When we remove the phenomenological damping term, the simulations reveal an interesting effect at high rotation frequencies and measurement times well beyond the $T_2$ of the diamond sample used in our experiments. For rotation speeds $<200\,000\,$rpm, we observe the familiar rapid damping of contrast at non-zero $\theta_B$ as the rotation speed increases. However, as we exceed rotation speeds possible in our current experiments, a revival of spin-echo contrast at twice-integer periods of the rotation frequency appears for small tilt angles. This behavior persists even after averaging over multiple baths to simulate measurement of an ensemble of NV centers. Figure \ref{fig:fig4} (a) shows the spin-echo signal as a function of time, evaluated at integer multiples of $4\pi/\omega_{13}$ (\emph{i.e.} at each $^{13}$C revival), for various rotation speeds at $\theta_B = 15^\circ$. In Fig. \ref{fig:fig4}(b) we examine the onset of rotational modulation as the magnetic field tilt angle is increased at rotation speeds of 200,000\,rpm (3.333\,kHz) and 500,000\,rpm (8.33\,kHz).

The underlying cause of the initial spin-echo modulation at low magnetic field angles can be easily understood as arising from modulation of the NV $m_S = 0\rightarrow-1$ transition energy due to the angular dependence of the hyperfine interaction. This modulation is a small perturbation to the NV and nuclear spin eigenstates at the rotation frequency $\omega_\text{rot}$. When measured phase-incoherently due to the random positions of $^{13}$C nuclei, the modulation leads to revivals in the spin-echo signal, in direct analogy and on top of the original revivals at the $^{13}$C Larmor period. As the tilt angle is increased, the degree of state mixing between electron spin levels and the consequent electron-spin augmentation of the nuclear spin precession begins to become significant. This process suppresses coherent revivals of the spin echo signal due to the strongly perturbed nuclear spin precession rate. 

Beyond the initial collapse of the spin-echo contrast with tilt angle (vertical axis of Fig. \ref{fig:fig4}), the simulation exhibits several low-amplitude contrast revivals (particularly evident for the faster rotation speed in Fig. \ref{fig:fig4}(b)). We attribute these effects to insufficient averaging over different spin bath configurations in our simulations, rather than any possible coherent effect induced by rotational averaging. We note that when the dipolar coupling term in our simulations is set to zero, these features remain, suggesting they originate from particular spin baths (\emph{i.e.} no strongly coupled nuclear spins). We would not expect to detect these in any experiment with an ensemble of NV centers, but leave open the possibility that single NVs with a few nearby $^{13}$C spins may exhibit similar effects.

\section{Discussion}\label{sec:disc}
The spatial variation of the NV-$^{13}$C hyperfine interaction in the presence of an off-axis field can be best described as each nuclear spin precessing at a rate determined by its position relative to the NV. When rotated, the nuclear spins precess around an axis determined by the external magnetic field and the local NV-dipole field. Physical rotation imparts a relative rotation between axes within and external to the diamond, in this case the NV axis and the nuclear precession axis. Therefore, the nuclear precession frequency becomes a function of time. This can be intuitively understood as frequency-modulation of the nuclear spin precession with a modulation depth determined by the position of the $^{13}$C nucleus. 

For practical purposes, any effects of rotation on the weak intra-bath homonuclear couplings with the magnetic field at or near the magic angle is totally obscured by the effects of the modulated NV-$^{13}$C hyperfine interaction. The theoretical model presented in the previous section corroborates this reasoning and predicts that at small tilt angles, the NV spin echo signal is modulated with contrast revivals at integer multiples of twice the rotation period. Experimental limitations prevent us from observing this directly, but no evidence of significant NV-$^{13}$C decoupling is expected given the small angles where contrast remains visible. To progress towards homonuclear decoupling with rotation, strategies to mitigate the NV-nuclear spin interaction must first be investigated. We discuss one such strategy, based on slow rotation, below. 

While sample rotation is a ubiquitous tool in NMR, the considerably stronger electron-nuclear spin couplings in EPR systems mean that mechanical rotation at rates comparable to the anisotropic interaction strength is impractical. Our example system offers a particularly stark example, as the peak homonuclear $^{13}$C-$^{13}$C dipolar interactions $(10^2-10^3$\,Hz) are easily reached in our rotating experiments, in comparison to the typical NV-$^{13}$C coupling ($10^5$\,Hz, 6\,million rpm). However, it is not strictly necessary to rotate faster than the anisotropic interaction strength in order to eliminate it. If the system is subjected to phase evolution under three equally spaced azimuthal angles (\emph{i.e.} $120^\circ$) of the magnetic field tilted by $54.7^\circ$ from the rotation axis, the anisotropic frequency shifts sum to zero. The phases accumulated at each angular position therefore sum in such a way as to yield only the isotropic component, a technique known as `magic angle hopping'~\cite{bax_chemical_1983, szeverenyi_magic-angle_1985}. 

Practically, the magnetic field reorientation is done by either `hopping' the sample rotor to each new position and performing a phase measurement of the spins, or rotating sufficiently slowly that the evolution is effectively static during the phase measurement. In either case, the result of the phase measurement is stored as a longitudinal spin state during the reorientation step, making the relevant time scale for rotation $\sim T_1$, which for many EPR systems, including the NV, can be on the order of rotation periods achievable with gas-powered magic-angle spinners~\cite{hubrich_magic-angle_1997, hessinger_magic-angle_2000}, or electric motors, as in our case. The NV-$^{13}$C system, with $T_1$ typically $>1\,$ms, presents an intriguing system in which magic-angle hopping could be investigated. At rotation speeds of just a few kHz, three appropriate quantum measurements could then be performed in sequence, with the resulting phase devoid of any contribution from the NV-$^{13}$C interaction. Remarkably, these speeds should also be sufficient to induce homonuclear decoupling within the bath itself. 

Further work is needed to refine the magic-angle hopping measurement procedure for the case of the NV before an experimental demonstration is possible. For example, spin-echo is no longer a suitable quantum measurement, since it measures phase differences across the interrogation times, instead, $T_2^\ast$-limited Ramsey interferometry is more appropriate. More pressingly, the large zero-field splitting of the spin-1 NV effectively freezes its orientation in the diamond lattice, and so it will not align along the magnetic field at each rotation angle like a nuclear spin (or unbound spin-$1/2$ electron) would as the magnetic field is rotated. This may require a large magnetic field or a tilt angle exceeding the magic angle to observe cancellation of the anisotropic interaction, though neither are available in our current experimental configuration. In any case, some evidence of rotational averaging should be detectable even with imperfect cancellation, and this will be the subject of future work.

\section{Conclusions}

We have examined the prospects of using physical rotation of a quantum system for decoupling of deleterious spin bath interactions. We found that NV spin coherence is rapidly lost as the magnetic field is tilted away from the rotation axis, an observation we attribute to angular variation of the NV-$^{13}$C dipolar interaction. We have used theoretical models to support our experimental findings and predict new phenomenology that could be explored in an optimized experiment and different diamond sample. Our key finding is that at the rotation speeds possible in our experiment, the NV-$^{13}$C hyperfine coupling is made time dependent, and induces dephasing that swamps any observation of weaker homonuclear decoupling of the spin bath. Magic-angle hopping may be a feasible alternative to our experiments, allowing the electron-nuclear hyperfine interaction itself to be rotationally averaged. 

Unlike many other quantum sensing and QIP platforms, the NV center in diamond naturally lends itself to applications where motional and rotational degrees of freedom are not fixed~\cite{delord_electron_2017, delord_ramsey_2018, russell_optimising_2021}. In these situations, rotation of the diamond can be essential or detrimental to the intended application. The results presented in this paper form a crucial first step of characterisation, and lay the foundations for employing rotational averaging to decouple interactions between the central spin and the nuclear spin bath in solid state quantum sensing. Anisotropic magnetic dipolar interactions are becoming ubiquitous in emerging applications of quantum systems. Novel characterisation and control schemes based on rotation could allow for interesting new avenues of investigation in quantum sensing and information processing, as well as more fundamental directions such as many-body physics with strongly interacting solid state systems~\cite{kucsko_critical_2018, choi_observation_2017}.

\section*{Acknowledgements}
We thank L. T. Hall for fruitful discussions and comments on the manuscript. This work was supported by the Australian Research Council Discovery Scheme (DP190100949).

\end{document}